\documentclass[12pt,fleqn]{article}

\usepackage{epsfig}

\newcount\xrefpos \xrefpos=0
\newcount\yrefpos \yrefpos=0
\newcount\xput \xput=0
\newcount\yput \yput=0

\def\refpos#1 #2 #3{\global\xrefpos=#1 \global\yrefpos=#2
                         \rlap{$\smash{#3}$}}
\def\put #1 #2 #3{\xput=#1 \yput=#2
                  \advance\xput by -\xrefpos
                  \advance\yput by -\yrefpos
                  \rlap{\kern\the\xput truebp
                        \vbox to 0pt{\vss\hbox{$\displaystyle #3$}
                        \kern\the\yput truebp}}}
\def\beginlabels\refpos#1\endlabels{\hbox{$\refpos#1$}}

\newcommand{\lb}{\langle}
\newcommand{\rb}{\rangle}
\newcommand{\sla}{\!\!\!/}

\newcommand{\eps}{\epsilon}

\newcommand{\A}{\ensuremath{{\cal A}}}

\newcommand{\be}{\begin{equation}}
\newcommand{\ee}{\end{equation}}
\newcommand{\beq}{\begin{equation*}}
\newcommand{\eeq}{\end{equation}}
\newcommand{\bea}{\begin{eqnarray}}
\newcommand{\eea}{\end{eqnarray}}

\begin{document}

\begin{titlepage}

\bigskip
\hskip 4.8in\vbox{\baselineskip12pt \hbox{hep-ph/0510148}}

\bigskip
\bigskip
\bigskip

\begin{center}

{\Large \bf  Recursion relations, Helicity Amplitudes and
Dimensional Regularization}
\end{center}

\bigskip
\bigskip
\bigskip

\centerline{\bf Callum Quigley$^1$   and Moshe Rozali$^2$ }

\bigskip
\bigskip
\bigskip

\centerline{ \it $^1$Department of Mathematics} \centerline{\it
University of Toronto} \centerline{\it Toronto, Ontario M5S 2E4,
Canada} \centerline{\small \tt cquigley@math.utoronto.ca}
\centerline{}

\centerline{ \it $^2$Department of Physics and Astronomy}
\centerline{\it University of British Columbia} \centerline{\it
Vancouver, British Columbia V6T 1Z1, Canada} \centerline{\small \tt
rozali@phas.ubc.ca}

\bigskip
\bigskip

\begin{abstract}
\vskip 2pt Using the method of on-shell recursion relations we
compute tree level amplitudes including D-dimensional scalars and
fermions. These tree level amplitudes are needed for calculations of
one-loop amplitudes in QCD involving external quarks and gluons.
\end{abstract}

\end{titlepage}

%\newpage

\baselineskip=18pt \setcounter{footnote}{0}

\section{Introduction and Summary}

The last two years have seen the development of new and surprising
techniques for  performing perturbative calculations in gauge
theories. Following the seminal paper by Witten \cite{Witten}
(drawing on earlier insights by Nair \cite{Nair}), quantifying  and
generalizing the simplicity of some tree level amplitudes \cite{PT},
the initial efforts have been focused on the relation to twistor
space (for a review see \cite{twistor}). These investigations
resulted in a beautiful effective Feynman diagram technique
\cite{csw}, the so-called MHV diagrams or CSW rules, which can be
applied to calculations of tree level \cite{tree}, one-loop
\cite{loop} amplitudes and amplitudes with external massive sources
\cite{massive1}. Recently it has been extended to gravity amplitudes
as well \cite{Bjerrum-Bohr:2005jr}.

Though the CSW diagrammatic method proved much more efficient in
calculating amplitudes than the traditional method of Feynman
diagrams, they still have the same flavor,  and it turns out there
exists an even more efficient method, that of the on-shell recursion
relation. This method was suggested in \cite{BCF} (see also
\cite{diss}), and was proven by \cite{BCFW}. It seems applicable to
calculations of tree level amplitude (or more generally rational
functions) in a wide range of theories, and has the flavor of the
analytic S-matrix theory, in that it does not make use of  off-shell
structures. As such it seems like a genuinely new way of performing
calculations in perturbative quantum field theories.

The recursion relations are reviewed in section 2 below, together
with other background material. They have been utilized to calculate
tree level amplitudes in gauge theories \cite{treebcf}, gravity
\cite{gravitybcf}, amplitudes including massive sources
\cite{massbcf,massbcf2,massmore} and rational functions appearing in
one-loop amplitudes \cite{rational}. In all these cases the results
obtained are either new, or are a more compact form of previously
calculated results. The relation to the CSW method was discussed
recently in \cite{Risager:2005vk}.

A more challenging task, and one of relevance to upcoming
experiments at the LHC, is the calculation of one-loop amplitudes.
The main ingredient used in calculating one loop amplitude is that
of unitarity: the multi-valued part of the loop amplitude is
determined by the tree level results (see for example
\cite{Bern:1994zx,Dixon:1996wi}). The compactness of the results for
the tree level amplitudes is extremely useful when they are used as
input for the calculation of one-loop amplitudes.

 The method of generalized unitarity is one of the most efficient general methods of using
the knowledge  of  tree level amplitudes in calculating the one-loop
amplitudes. It has been discussed recently in \cite{gen,pure}. In
particular then discussion in \cite{gen} has concentrated on
cut-constructible amplitudes (in the sense of  \cite{fusing}). The
more general amplitude has rational pieces, which can sometimes be
determined separately using recursion relations \cite{boot}.

However, as explained in \cite{pure}, a systematic method to obtain
the complete amplitude, including any rational parts, is using
generalized unitarity in D-dimensions. In continuing away from four
dimensions, in dimensional regularization, the rational pieces
acquire cuts as well, and therefore can be constructed using
generalized unitarity. The form of the one-loop amplitude thus
constructed is expected to be  compact, reflecting the simplicity of
their building blocks, the tree level diagrams.

Motivated by this line of development, we calculate below all the
tree level amplitudes needed for calculating one-loop amplitudes
with up to five partons.  These tree level amplitudes differ from
the ones previously calculated in that some of the external legs are
continued to D-dimensions. The one loop amplitudes with  5 partons
were previously  calculated in \cite{five}, and we can check the
generalized unitarity method against those explicit results.
Calculations of one loop amplitudes in QCD, with up to six partons,
are in progress \cite{progress}.

The outline of this paper is as follows: in section 2 we motivate
the set of tree level amplitudes constructed here, as the building
blocks for the aforementioned one-loop amplitudes. The method we use
is the  extension of  the BCF recursion relations, and we explain
the new issues arising when including D-dimensional scalar and
fermions. In section 4 we exemplify the method by calculating the
four point amplitudes in detail. Section 5 is devoted to
calculations of the five point amplitudes and some checks on them.

\section{Preliminaries}
\subsection{Notations}

A massless momentum in four dimension can be written as a product of
two (bosonic) spinors which we denote by $\lambda_\alpha$ and
$\tilde{\lambda}_{\dot{\alpha}}$, so that $p_{\alpha\dot{\alpha}} =
\sigma^\mu_{\alpha\dot{\alpha}}p_\mu=\lambda_\alpha\tilde{\lambda}_{\dot{\alpha}}$,
where $\sigma^\mu= (1,\vec{\sigma})$, and $\vec{\sigma}$ are the
Pauli matrices. We also denote alternatively $\lambda_\alpha=
|\lambda \rb$ and $\tilde{\lambda}_{\dot{\alpha}}= [\lambda|$, so that $p=|\lambda\rb [\lambda|$. In
the case of several momenta, we also shorten $| \lambda_i \rb$ and
$[\lambda_i |$ to $  |i \rb$ and $[i|$ respectively.

 For four dimensional fermions we denote by $u_{\pm}(k)$ two of the
 positive energy solutions of the massless Dirac equation $k\!\!\!/u_{\pm}(k)=0$ of helicities $\pm \frac{1}{2}$.
 These solutions are denoted   $u_+(k) = |\lambda \rb = \lambda_\alpha$
 and $u_-(k)= |\lambda]=\lambda^{\dot{\alpha}} $, and they are eigenspinors of
 $\gamma^5$   with  eigenvalues $\mp$ respectively.

 The internal products of these spinors are
 defined as \be \lb ij \rb = \bar{u}_-(k_i) u_+(k_j) ~~~~~~~~~[ij]=
 \bar{u}_+(k_i) u_-(k_j) \ee which also defines the dual (bra) of each
 spinor (ket). With this notations $|i]=\lambda_i^{\dot{\alpha}}$ and
 $[ i|= (\lambda_i)_{\dot{\alpha}}$ are positive
 chirality spinors of opposite ($\mp$) helicities, whereas $|i\rb=(\lambda_i)_\alpha$ and $\lb i| =\lambda_i^\alpha$
 are negative chirality spinors of opposite ($\pm$) helicities (note that the internal products are then Lorentz scalars). These
 consist of the four positive energy solutions of the massless Dirac
 equation. One has   the following identities \be |i \rb
 [ i| =\omega_+\,  k_i \!\!\!/
  ~~~~~~~~~~~~~~~~~~~~~~~~~~~|i]\lb i|=\omega_-  \, k_i \!\!\!/
 \ee where $\omega_{\pm} = \frac{1}{2} (1 \pm \gamma^5)$ are
 projections onto the positive or negative chirality subspaces.

   Wave-functions for external gluons can be
 written in this basis as bi-spinors $\epsilon^{\pm}_{\alpha \dot{\alpha}}$ \be \epsilon^+(k) =  \, \frac{|q \rb  [k|
  }{\sqrt{2} \lb qk\rb} ~~~~~~~~\epsilon^-(k) = - \,
 \frac{ |k \rb  [q|
 }{\sqrt{2}[ qk ]} \ee where $q$ is an arbitrary reference
 momentum, changing it amounts to a gauge transformation on the
 gluon polarization vector. Similar expression hold for the conjugate part
 of the polarization $(\epsilon^{\pm})^{\dot{\alpha} \alpha }$. Working with
 $\epsilon^{\pm}_{\alpha \dot{\alpha}}$ amounts to concentrating on a negative helicity Weyl spinor, which does not mix with the positive helicity one in a purely massless theory.

 In performing helicity amplitude calculations, it is customary to
 include the wave-functions for internal fermions or gluons in the
 interaction vertices rather than in the propagator. Therefore the
 propagator is always the scalar propagator $\frac{1}{p^2}$, and
  numerators  which usually accompany fermion
  propagators  come
  about
 from the two interaction vertices connected by the given
 propagator\footnote{For fermions in a complex representation of the gauge group  there is a distinction
 between fermion and anti-fermion. As the formalism treats them uniformly, one has to add by hand
 a minus sign for each anti-fermion line.}.

 For other conventions and notation used in helicity amplitude
 calculations, we refer the reader to the review
 \cite{Dixon:1996wi}.

\subsection{One-Loop Amplitudes with Quarks and Gluons}
The tree level amplitudes calculated here are to be used as basic
building blocks in evaluating one-loop diagrams in pure QCD, with
quark-antiquark pair, and a number of external gluons. We briefly
describe here this calculation, using generalized unitarity in
$D=4-2\epsilon$ dimension, leaving the details to future work
\cite{progress}. This serves as a motivation for the particular set
of tree level amplitudes we evaluate in this paper.

We will concentrate on color-ordered (or partial) amplitudes,
stripping the color indices off the external legs. We will therefore
not distinguish between various matter representations\footnote{For
example,  we will not assume the fermions to be adjacent, as could
be assumed  for quarks, to include the possibility of external
gluinos.}. For the purposes of QCD, the fermions are in the
fundamental representation, and discussion of the color ordered
amplitudes in this context can be found e.g in \cite{multi}.

Another standard tool is the supersymmetric decomposition of
amplitudes. The one-loop amplitudes are easier to calculate if a
complete supersymmetric multiplet   runs in the loop, as loop
amplitudes are cut constructible in those cases \cite{fusing}. This
allows us to trade some combinations of particles running in the
loop for others. For our case, it is sufficient then to calculate
loop amplitudes with scalars and fermions (but no gluons) appearing
in the loop, and (as mentioned above) external quarks and gluons.

The one loop amplitudes can be calculated if we know their
singularities in $D=4-2\epsilon$ dimensions. This requires knowledge
of tree amplitudes with two external legs continues to $D$
dimensions. Using four dimensional helicity regularization (i.e,
where the momenta but not the polarizations are continued to $D$
dimensions), the $D$ dimensional external momenta can be thought of
as 4 dimensional massive momenta \cite{massive}. Inspection of the
quadruple and triple cuts of the aforementioned one-loop amplitudes
leads to the following 4 sets of tree level amplitudes (see figure
1), which will be calculated below:

\begin{itemize}
\item Amplitudes including  2 massive (or massless in $D=4-2\epsilon$ dimensions)
scalars, and some number of gluons. These   were already calculated
in the first reference in \cite{massbcf}.

\item Amplitudes including 2 massive fermions (which we denote by
$\lambda$), and some external gluons. We will label these type-A
amplitudes.
\item Amplitudes including 2 massive scalars, and massless gluons,
accompanied by an external quark-antiquark pair.  We will label
these type-B amplitudes.

\item Amplitudes including a massive scalar and massive fermions,
with external massless fermion and  a few  gluons. Those amplitudes
will be labeled type-C.
\end{itemize}
\begin{figure}[h]
\centerline{
         \psfig{figure=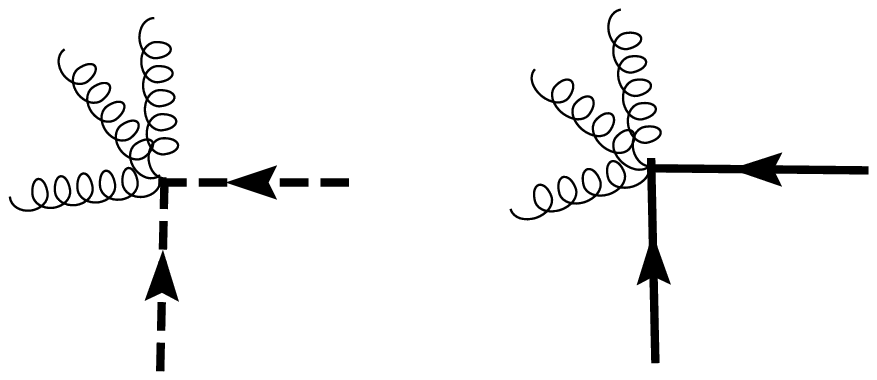,height=1in}
         \hskip 1cm \psfig{figure=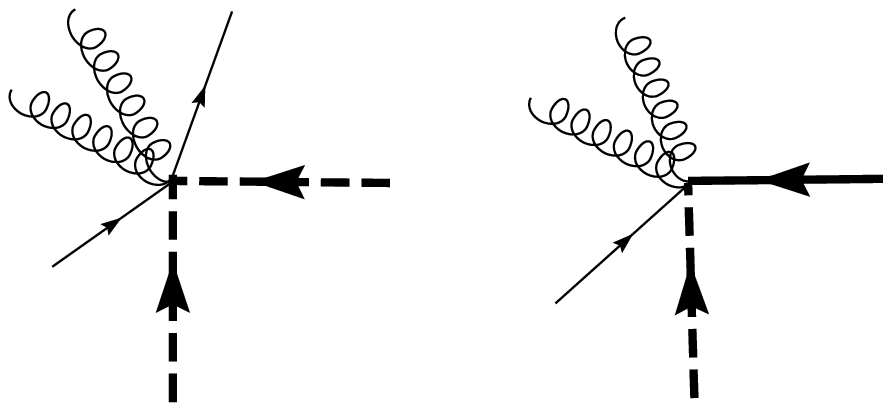,height=1in}}
         \caption{The relevant tree diagrams, with varying number of gluons, are ordered from left to
         right. The massive legs are denoted in boldface lines, solid lines denote fermions, dashed lines
         scalars, and wiggly lines gluons.}
\end{figure}
We note that all amplitudes involve two massive legs which are
adjacent.

In an effort to use  a uniform notation, we will denote the momenta
of external\footnote{We will not use uniform notation for
intermediate state momenta that are eliminated from the final result
anyhow.} fermions by $k$, external scalars by $l$, and external
gluon by $p$. In addition, as explained in the next subsection we
need to distinguish between four dimensional momenta (denoted by
small letters) and D-dimensional ones (denoted by capital letters).

\subsection{D-dimensional Fields}
We find that in order to use D-dimensional unitarity, we have to use
helicity methods in calculating tree level amplitudes in
D-dimensions. In the FDH regularization scheme, the momentum is
continued to D-dimension. Every D-dimensional momentum $P$ can be
decomposed as $P=p+\mu$, where $p$ is the four dimensional
component, and $\mu$ is a component in a formal
$(-2\epsilon)$-dimensional orthogonal space\footnote{We will use the
capital letters for D-dimensional momenta, and small letters for
their four dimensional components. The $(-2\epsilon)$ component of
all D-dimensional momenta is always $\pm \mu$.}. Working in mostly
minus signature, $P^2 =p^2-\mu^2$, so on shell massless momentum
($P^2=0$) is equivalent to four-dimensional massive momentum
$p^2=\mu^2$. Therefore for scalars, working away from 4 dimensions
is equivalent to adding mass to the scalar field.

Each loop momentum integration can be decomposed as $\frac{d^D
P}{(2\pi)^D} = \frac{d^4 p}{(2\pi)^4}\frac{d^{ (-2\epsilon)
}\mu}{(2\pi)^4}$, so the mass $\mu$ is always integrated over, and
the $\epsilon$ dependence of the amplitude is generated from the
$\mu$ dependence of the integrand. Similarly, each vertex is
accompanied by a delta function imposing momentum conservation. As
all momenta are now D-dimensional, those are also D-dimensional
delta functions. It is therefore necessary (in general)  to regard
the momenta of internal lines as being D-dimensional as well; for
tree level amplitudes they are just given as linear combinations of
external momenta, given by imposing all the momentum conservation
constraints.

For internal fermionic lines one always has to sum over the intermediate
spinor wavefunctions, so choice of basis is not necessary. We will
therefore use the notation $|P\}$, $\{P|$ to refer collectively to these wavefunctions. The sum over
the intermediate wavefunction is performed using the identity \be
|P\}\{P| = P \!\!\!/ \ee Note that now $ P \!\!\!/$ has one component ($ p \!\!\!/$) that preserves helicity  and one that flips it($ \mu \!\!\!/$). similarly,    in D-dimensions the components $p\!\!\!/$ and $\mu\!\!\!/$ behave
differently with respect to chirality, $\{p\!\!\!/, \gamma^5 \}=0$
whereas $[\mu\!\!\!/, \gamma^5 ]=0$.

To mimic the four dimensional helicity methods, we want to utilize helicity-like states for external fermions, even when they are D-dimensional. As we keep $\gamma^5$ four dimensional, we can still
use chiral basis which we denote as before by $|P\rb = \omega_+ |P\}$ and $|P]= \omega_- |P\}$ (and similarly
the conjugates $\lb P|$ and $[P|$).
However the states of definite helicity, $|P\rb$ and $[P|$ (or similarly   $|P]$  and $ \lb P$) now mix with each other.
 Indeed, the basis vectors $|P\rb$ and $|P]$ do not individually satisfy the massless Dirac equation in D-dimensions. That equation written in terms of these Weyl fermions is,
 \be p\!\!\!/|p\rb +\mu\!\!\!/|p]=0~~~~~~~~
p\!\!\!/|p] +\mu\!\!\!/|p \rb=0 \ee which is consistent with the
mass-shell condition $p^2 =\mu^2$.

Nevertheless one can assemble the physical amplitudes (with external wave functions $|P\}=|P\rb +|P]$) from the ones calculated here, as helicity violation is limited to insertions of $\mu\!\!\!/$ in fermion lines. In the course of using the recursion relations we demonstrate this process, which is also relevant for the calculation of one-loop amplitudes using generalized unitarity \cite{progress}. Whenever one encounters an intermediate D-dimensional fermion, one can write the numerator of the propagator as
\be P \!\!\!/= |P\}\{P|   = (|P\rb +|P]) (\lb P| +[P|)\ee
and make use of the partial amplitudes with helicity    states, the ones we calculate here.

Note however that the propagator has both helicity  preserving and helicity  flipping parts. The helicity  preserving parts are the usual
propagators, usually drawn as connecting $\pm$ states to $\mp$ states,
 \be |P \rb
 [ P| =\omega_+\,  p \!\!\!/
  ~~~~~~~~~~~~~~~~~~~~~~~~~~~|P]\lb P|=\omega_-  \, p \!\!\!/
 \ee
  whereas the helicity  flipping parts are new, and   are the part of the propagator that connects $\pm$ states to $\pm$ at the other end of the propagator. They arise from the identities \be
   |P \rb \lb P| =\omega_-\,  \mu \!\!\!/ ~~~~~~~~~~~~~~~~~~~~~~~~~~~
 |P][ P| =\omega_+
 \,  \mu \!\!\!/ \ee

 The main advantage of using   chiral external states is that one can use the  simple   expressions for the gluon polarizations, \bea
 \epsilon\!\!\!/^+(k) &=&  \, \frac{1}{\sqrt{2} \lb qk\rb} \left(|q\rb  [ k| +  |k] \lb q |\right)= (\epsilon^+)_{\alpha \dot{\alpha}} +(\epsilon^+)^{\dot{\alpha}\alpha }\nonumber \\
  \epsilon\!\!\!/^-(k) &=& - \,
 \frac{1}{\sqrt{2}[ qk ]}  \left(| q]\lb k| + |k\rb [q|\right)=(\epsilon^-)_{\alpha \dot{\alpha}} +(\epsilon^+)^{\dot{\alpha}\alpha }
  \eea
  This polarization contracts into the fermionic states that accompany the gluon in an interaction vertex. In case at least one of these states is a Weyl fermion, one of the terms in the polarization vanishes (which one depends on the chirality of the fermionic state). Note that in this case the chirality of the other fermion is determined, even if it is a D-dimensional (and thus Dirac) fermion.

  To summarize the step of the calculations, we will be using the BCF
recursion relation to find an expression for helicity amplitude
which is valid in general dimension D (in the FDH scheme).
Subsequent manipulations will depend on which external (and
internal) legs are taken to be D-dimensional, those include
translating the helicity states into momenta, using mass shell
conditions and trace identities to simplify the results. We will be
very explicit in calculations of the four point amplitudes in
section 3, to demonstrate the issues involved, and will be less
detailed in deriving the five point amplitudes.

\subsection{Recursion Relations}

To evaluate the tree level amplitudes, we will use the recursion
relations first discussed in \cite{BCF} and proven in \cite{BCFW}.
The proof utilizes the analytic properties of rational functions,
and can be then generalized to massive particles \cite{massive}, and
to purely rational loop amplitudes \cite{loop}, or to calcualting
the rational parts of loop amplitudes \cite{rational}. Here we
slightly generalize it for the case of D-dimensional fields.  We
briefly review the method as needed for our purposes, highlighting
the slight differences arising in our case. We refer the reader to a
more detailed discussion in the original papers.

We will be discussing tree level amplitudes $\A_n(p_1, ...,p_n)$ of
$n$ external on-shell particles, $n-2$ of which are massless  in 4
dimensions, or two are taken to be  massless in $D=4-2\epsilon$
dimensions (or equivalently massive in 4 dimensions). The recursion
relations depend on choosing two of the external momenta (labeled
$i,j$) and "marking" them.

Now, define a function $\A(z)$ to be the amplitude evaluated at the
shifted momenta\footnote{Note that the momentum conservation
constraint is unaltered by the shift, so $\A(z)$ can be calculated
using perturbation theory.}
 $\hat{p_i}=p_i+z\eta$ and
$\hat{p_j}=p_j-z\eta$, where $\eta $ is a null vector orthogonal to
both $p_i,p_j$. This ensures the same mass-shell condition applies
to the shifted momenta. The shifted momenta are now null and
complex. We will always choose at least one of the marked momenta to
be massless in 4 dimensions, as this simplifies the analysis, and is
sufficient for our purposes\footnote{For simplicity we take $\eta$
to be purely four dimensional null momentum, even when D-dimensional
momenta are involved.}. In case both marked momenta are null, we can
write them as product of spinors: $p_i=\lambda_i \tilde{\lambda_i}$
and $p_j=\lambda_j \tilde{\lambda_j}$, then $\eta=
\lambda_j\tilde{\lambda_i}$. In this case the shift amount to
shifting the spinors $\lambda_i \rightarrow \lambda_i +z \lambda_j$
and $\tilde{\lambda_j} \rightarrow \tilde{\lambda_j} -z
\tilde{\lambda_i}$, leaving $\tilde{\lambda_i}$ and $\lambda_j$
intact.   In case that $p_i$ is massive and $p_j = |j\rb [j|$ is
massless, we have $\eta = [j|p_i|j\rb$ (and similarly for
$i\leftrightarrow j$).

 One then divides the $n$ external momenta to
two cyclically ordered groups, which are labeled
$L=\{p_r,...p_i,...,p_s\}, R=\{p_{s+1},...p_j,...,p_{r-1}\}$. As is
indicated the groupings is such that $p_i\,\epsilon\, L$ and
$p_j\,\epsilon\, R$, and we will sum over all such groupings. We
denote by $p= p_r+...+p_s$, the momentum  flowing in the channel
between the $L,R$ groups of momenta\footnote{ To conform with the
notation in this paper, we use $p$ for four dimensional intermediate
momentum, and $P$ for a D-dimensional such momentum.}, and $\hat{p}
= p+ z\eta$ is the shifted intermediate momentum. The shift variable
$z$ is chosen to impose the appropriate mass-shell condition on the
shifted intermediate momentum $\hat{p}$. For uniformity of notation,
we impose the same mass-shell condition imposed on external legs:
$\hat{p}^2=0$ for purely four dimensional momenta, and
$\hat{p}^2=\mu^2$ for components of D-dimensional momentum
$\hat{P}$.

The BCF recursion relation is then \be \A_n(p_1,...p_n)=\sum_{L,R}
\sum_h A_L(p_r,...\hat{p_i},...,p_s, -\hat{p}^h) \frac{1}{p^2}
A_R(\hat{p}^{-h},p_{s+1},...,\hat{p_j},...,p_{r-1}) \ee The first
sum is over all possible groupings of external momenta, as described
above. The second sum is over all possible intermediate states
(depending on the matter content of the theory), and their
helicities\footnote{For D-dimensional momenta helicity is not
well-defined, but one still have to sum over an appropriately chosen
basis, as described above.} $h$. The amplitudes $A_L,A_R$ depend on
shifted momenta as indicated, and are momentum conserving on-shell
amplitudes, albeit with complex momentum; they include an additional
external leg with momentum $\pm \hat{p}$ and the appropriate
helicity $\pm h$. Note also that the momentum $p$ appearing in the
propagator is unshifted.

The validity of the recursion relations depends crucially on one
technical assumption, that of the vanishing of the function $\A(z)$
as $z\rightarrow \infty$. This depends on the helicities of the
marked momenta $i,j$. In general, the helicities $(h_i,h_j)$ must
chosen so that $h_i\geq h_j$, with additional constraints when
choosing quarks. In particular, for two gluons we cannot choose
$(-,+)$ \cite{BCF}; when choosing a gluon and a scalar, positive
helicity gluons must be particle $i$, while negative helicity gluons
must be in the position $j$ \cite{massbcf}. When ``marking" a quark,
we cannot also choose an adjacent quark nor an adjacent scalar, and
for an adjacent gluon-quark pair, we have similar rule as above:
positive and negative helicity gluons must be chosen as $i$ and $j$,
respectively \cite{massbcf2}.

We would now like to resolve the issue of choosing an adjacent gluon
and quark (in that order) with helicities $(+,+)$. While
\cite{treebcf} claims that this choice is invalid, we will argue
that it is in fact allowed, as stated in \cite{massbcf2}. We will
consider $\A(z)$ as a sum of Feynman diagrams where the momenta of
the gluon $i$ and adjacent quark $j$ depend on $z$, and we will
write $\hat{p}_i=p(z)$ and $\hat{p}_j=k(z)$. When $i$ and $j$ share
a vertex, the $z$-dependance of those diagrams are completely
determined by the corresponding polarization vector and
wavefunction. This class of diagrams contributes the factor
$[k(z)|\left(\eps^+\left(p(z)\right)\right)$. We recall that
$[k(z)|\propto z$ and $\eps^+\left(p(z)\right)\propto 1/z$, and so
naively this term would not decay for large $z$. However, we point
out that \be \qquad [k(z)|\eps^+\left(p(z)\right)=\frac{[k(z)\ p]\lb
q|}{\lb q\ p(z)\rb}=\frac{[k\ p]\lb q|}{\lb q\ p(z)\rb},\ee so this
type of diagram does vanish for large values of $z$. When the
particles $i$ and $j$ are separated by any single ($z$-dependent)
propagator, it is straightforward to check that similar cancelations
occur in the numerator whenever an amplitude poses the threat of not
decaying at large $z$. Finally, any further lengthening of the
``$z$-path" is harmless since the only components which grow with
$z$ are the scalar-gluon and the triple-gluon vertices. However,
introducing such interactions must be accompanied by the appropriate
bosonic propagator, which then compensates for the $z$ behaviour of
the vertex. Thus, we find that the choice $(+,+)$ \textit{is} valid
for an adjacent gluon-fermion pair (in that order), and similarly
for the choice $(-,-)$.

 The new element in our calculation is  the continuation away from four dimensions.
 However, the continuation to $D$ dimensions does not affect the
analytic  properties of $\A(z)$ or any of its ingredients, so the
the recursion relations remain valid, as do the choices for $(i,j)$
given in the literature. This fact was already noted in
\cite{massbcf2}. We point out that the only qualitatively new
ingredient in $D$ dimensions is the helicity-flipping fermionic
propagator. However, this in fact has even better large $z$
behaviour then its standard helicity-preserving counterpart, since
the numerator $\mu$ is constant, as opposed to $\hat{P}\sla$ which
is linear in $z$.

To summarize, in the following we will use the BCF recursion
relations, always marking momenta in configurations that are proven
to be allowed (that is, when the vanishing of the boundary term has
been established, as described above).

%$(-,+)$, but (with a few exceptions we will explain momentarily) one
%can show that the vanishing at infinity holds otherwise. The
%additional constraints we impose are that if two fermions or a
%conditions for the validity of the recursion relations are
%independent of the continuation to D-dimensions, therefore we will
%be able to use those marked momentum configurations for which the
%validity was already proven.

\subsection{Primitive Vertices}

 We are interested in calculating tree level diagrams involving
 massless quarks and gluons, and massive  scalars and fermions.
 Using the BCF recursion relations \cite{BCF}, we need as basic
 building blocks a few cubic amplitudes. We list here the vertices
 we need for our calculation. In every case these are simply constructed from the
 gauge theory action by contraction with the external wavefunction
 of the appropriate helicity. In all the cases listed below we treat
 all momenta as incoming. All vertices are written for $(i,j,k)$ cyclically
 ordered, the expressions for the other cyclic ordering
$(j,i,k)$ differ by an overall sign   if $(i,j)$ are fermions.

 The primitive vertices we need are:
\begin{itemize}
\item  gluon-fermion vertex (figure 2):
\bea \mbox {negative helicity gluon k:}&\,\,\, \frac{\lb ik \rb [ q
j ]}{[ q k ]}   \nonumber\\  \mbox {positive helicity gluon
k:}&\,\,\, -\frac{\lb iq \rb [ k j]}{\lb q k \rb} \eea where q ia an
arbitrary reference null vector with $q = |q\rb [q|$. The vertex is
independent of this choice as a consequence of gauge
invariance. As each one of the primitive vertices is
gauge invariant, these reference vectors can be chosen independently
for each vertex.
\begin{figure}[h]
\centerline{
         \psfig{figure=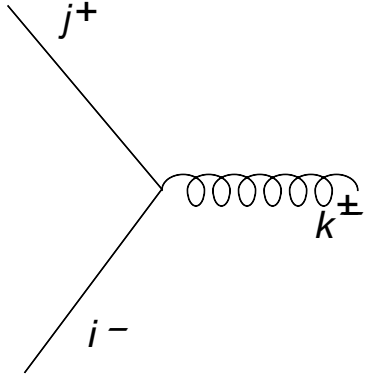,height=2in}}
         \caption{ Gluon-Fermion primitive vertex, the fermion helicities are drawn, the gluon
          can
         have positive or negative helicity.}
\end{figure}
\item scalar-fermion vertex (figure 3):
\bea \mbox {negative helicity external legs:}&\,\,\,  \lb ij \rb\nonumber\\
\mbox {positive helicity external legs:}&\,\,\, -[ij]\eea
\begin{figure}[h]
\centerline{
         \psfig{figure=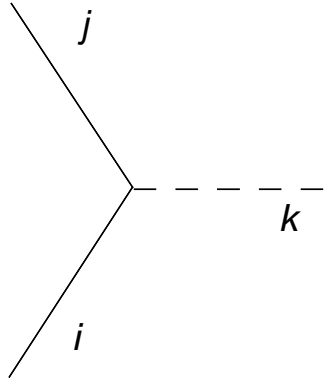,height=2in}}
         \caption{ Scalar-Fermion primitive vertex, The 3 external legs have the same
         helicities.}
\end{figure}
\item scalar-gluon vertex (figure 4):
\bea \mbox {negative helicity gluon k:}&\,\,\,   -\frac{  \lb k|j|q]}{[qk]} \nonumber\\
\mbox {positive helicity gluon k:}&\,\,\, \frac{\lb q|j|k] }{\lb q k
\rb} \eea
\begin{figure}[h]
\centerline{
         \psfig{figure=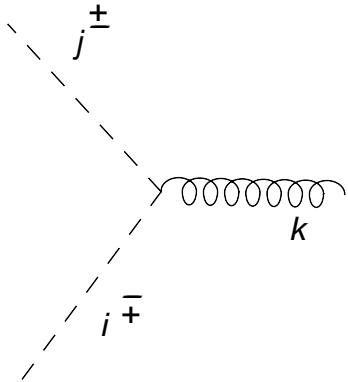,height=2in}}
         \caption{ Gluon-scalar primitive vertex, the scalar helicities are drawn, the gluon
          can
         have positive or negative helicity.}
\end{figure}
\item gluon cubic vertex (figure 5):
\bea \mbox {MHV vertex:}&\,\,\,  \frac{\lb ij \rb^3}{\lb jk \rb \lb ki \rb }\nonumber\\
\mbox { $\overline{MHV}$ vertex:}&\,\,\, \frac{[ ij ]^3}{[ jk ] [ ki
]}\eea
\begin{figure}[h]
\centerline{
         \psfig{figure=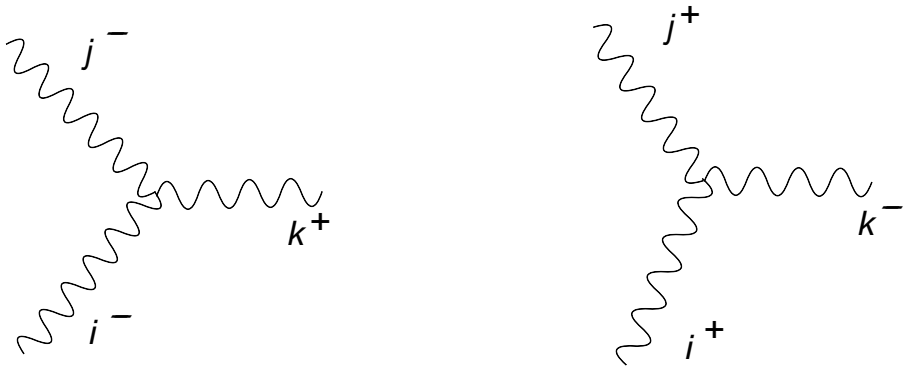,height=2in}}
         \caption{ Gluon MHV (left) and $\overline{MHV}$  (right) cubic vertices.}
\end{figure}
\end{itemize}
There are other non-vanishing cubic vertices which we will not need,
therefore we will not present them here. Additionally, for D-dimensional
fermions one utilizes the full expression for the gluon polarization, resulting in additional   terms in the interaction vertices which we write down when we use them below.

\subsection{Checks on the Amplitudes}

Though the on-shell recursion relations are proven, it is still
useful to perform a few checks on the resulting expression,
verifying
  the various intermediate steps leading to the final expression.
These steps include choosing marked momenta (such that $A(z)$
vanishes as $z \rightarrow \infty$), choices of various reference
momenta, and straightforward (but sometimes tedious) algebra.

The first check one can perform is comparison with the result of
Feynman diagram calculation. We have checked our expressions against
such calculations for all the four point amplitudes and some of the
five point ones. Typically the recursion relations yield much more
compact expressions for the amplitudes, so the main complication is
to reduce the complex Feynman diagram result to the simpler
expression.

Another check we have performed is (some of) the collinear limits of
the amplitudes. The collinear limits are a subset of the
multi-particle poles which occur at tree level amplitudes. As the
sum of two neighboring momenta becomes on-shell the amplitude
factorizes in the appropriate channel. All our amplitudes have poles
at the right location, and in some cases we have checked explicitly
that the residue of the pole is the expected one. We exemplify the
collinear limit in the appendix.

Finally,  in the limit $\mu \rightarrow 0$ all external legs are
four dimensional. In some cases the amplitudes are known in that
limit, and we reproduce these results.

%VANISHING AT INFINITY??? still need comments on marked moneta etc.

\section{Four Point Amplitudes}

 The amplitudes with four external
legs are fairly simple, and can be checked explicitly against
Feynman diagram calculations. We present the details and the results
in this section as a demonstration of the technique and the new
issues arising when D-dimensional fermions and scalars are included.

In addition, all such amplitudes can be seen to have the correct
factorization limits. Indeed, to see the singularity structure it is
 sufficient to inspect the denominators, we easily
see that they vanish if and only if the sum of two adjacent momenta
becomes on-shell (so spurious singularities are absent). Checking
factorization amount to verifying that the residue of these poles is
the expected one\footnote{The term collinear limit is inaccurate
when massive momenta are involved, what we really mean is
two-particle factorization limits.}.

\subsection{Type-A Amplitudes}
These amplitudes have  two adjacent massive fermions (of momenta and
helicities $K_1^+,K_2^-$) and two adjacent gluons (of momenta
$p_1,p_2$).  We discuss all helicity configurations in turn.

The first case of where the gluons are of opposite helicities  is
the amplitude  $\A_4(K_1^+,K_2^-,p_1^+,p_2^-)$. We choose the marked
momenta to be $(i,j)=(p_1^+,p_2^-)$, as this is one of the
configurations for which there is a general proof of vanishing at
infinity. In this case there is only one possible diagram appearing
in the recursion relation, with intermediate fermion of momentum
$P=K_2+p_1$. One gets   \be \A_4(K_1^+,K_2^-,1^+,2^-)= \frac{\lb K_2
q_1 \rb [\hat{1} -\hat{P}]}{\lb q_1 \hat{1} \rb} \,\frac{1}{P^2} \,
\frac{\lb \hat{P} \hat{2} \rb [q_2 K_1]}{[q_2 \hat{2}]} \ee where
$q_1,q_2$ are two reference momenta. Note that the helicity of the intermediate states of momentum $P$ is determined by that of the external momenta.

   We choose $q_1 = \hat{2}$ and
$q_2 =\hat{1}$, and use $|\hat{1}] =|1]$ and $|\hat{2}\rb =|2\rb$,
then this becomes \be~~~~~~~~~~~~~~~-\frac{\lb K_2 2\rb [1
|\hat{p}\!\!\!/ |2 \rb [1 K_1]}{\lb 2 \hat{1} \rb [1 \hat{2} ] P^2}
= -\frac{\lb K_2 2\rb [1 |k_2\!\!\!/ |2 \rb [1 K_1]}{\lb 2 1 \rb [1
2] P^2}\ee This result is valid in D-dimensions, with states such as
$|K_2\rb$  are defined in section 2. Note also that in this case the
relevant part of $P$ appearing in the intermediate state is $p$, as
$[1 |\mu \!\!\!/ |2 \rb=0$ by chirality selection rules.

The next step we simplify the result by taking the gluons to be
massless in four dimensions, and the fermion momenta to be massless
in D-dimensions (with $-2\epsilon$ component $\mu$). Then the four
dimensional component of $P$ is $p=p_1+k_2$, and the  $-2\epsilon$
component is $\mu$.   This result can be shown to be identical to
the one obtained from using Feynman graphs \be\label{fey}
\A_4(K_1^+,K_2^-,1^+,2^-)= \frac{(\epsilon_1^+ \cdot k_2) [K_1|\,
\epsilon_2^-\!\!\!\!/|K_2\rb}{(p_1+k_2)^2 -\mu^2} \ee  A slight
variation of the same calculation yields \be
\A_4(K_1^+,K_2^-,1^-,2^+)= - \frac{[K_1 2] \lb1 K_2\rb
[2|K_1\!\!\!\!/|1\rb}{P^2 \lb12\rb [21]}= \frac{(\epsilon_1^- \cdot k_1)
[K_1|\, \epsilon_2^+ \!\!\!\!/|K_2\rb}{(p_1+k_2)^2 -\mu^2} \ee where
the intermediate momentum $P=p_1+K_2$ in the only diagram
contributing to the recursion relations.

Additionally, for these gluon helicities there could be  one
helicity flipping amplitudes, using the same choices of marked and
reference momenta one gets: \be \A_4(K_1^+,K_2^+,1^+,2^-)=\frac{[
K_2
 1 ] \lb  2 \, -\hat{P}\rb}{\lb  2 1 \rb} \,\frac{1}{P^2} \,
\frac{\lb \hat{P}  2 \rb [ 1 K_1]}{ [1  2]}   \ee However, using $|P
\rb \lb P| =\omega_-\,  \mu \!\!\!/\,\,$  and $\lb 2|\mu
\!\!\!/|2\rb=0$ this amplitude vanishes.
%\be
%\A_4(K_1^+,K_2^+,1^+,2^-)=- \frac{[K_2 1][1 K_1] \lb 2|\mu
%\!\!\!/|2\rb }{\lb21\rb [12] \left[(p_1+k_2)^2-\mu^2\right]} \ee We
%note this is the helicity flipping part of the result (\ref{fey})
%computed using conventional Feynman diagrams.

Now, if the gluons are of the same helicity, we get the amplitudes
$\A_4(K_1,K_2,p_1^+,p_2^+)$, where the massive momenta $K_1, K_2$ are of either chirality. In this case both external states are of the form $|K_i\}=|K_i\rb+|K_i]$. Choosing the marked momenta to be
$(i,j)= (p_1^+,p_2^+)$ gives one possible intermediate
(D-dimensional fermionic) state with $P= K_2+p_1$.

When contracting with the gluon polarizations
$\epsilon\!\!\!/^+$ one gets the interaction vertex
\be \frac{1}{  \lb q \hat{1}\rb} \left(\lb \hat{P} q\rb [\hat{1} K_2] + [\hat{P} \hat{1}]
\lb q K_2\rb \right) \ee with a similar expression for the other
interaction vertex in the diagram. We get  \bea
\A_4(K_1,K_2,1^+,2^+)= \frac{1}{  \lb q_1 \hat{1}\rb} \left(\lb \hat{P} q_1 \rb [\hat{1} K_2] + [\hat{P} \hat{1}]
\lb q_1 K_2\rb \right) \, \frac{1}{P^2} \,\nonumber \\
  \frac{1}{  \lb q_2 \hat{2}\rb} \left(\lb -\hat{P} q_2\rb [\hat{2} K_1] + [-\hat{P} \hat{2}]
\lb q_2 K_1\rb \right)    \eea

There are four terms, corresponding to the four possible helicity assignments for $K_1, K_2$. As the amplitude vanishes in four dimensions, we know that only helicity flipping terms has to be retained.  Starting with $K_1^-,K_2^-$ we get
  (choosing $q_1=\hat{2}$ and $q_2= \hat{1}$),
 \be
\A_4(K_1^-,K_2^-,1^+,2^+)=  -\frac{ [\hat{P} \hat{1}]
\lb \hat{2} K_2\rb [\hat{P} \hat{2}] \lb \hat{1} K_1\rb}{   \lb 21\rb \lb 12 \rb}
   \ee
   using $|P ][ P| =\omega_+\,  \mu \!\!\!/\,\,$, and the more standard
   projections gives
 \be ~~~~~~~~~~~~~~~~~~~~~~~~~~~~  -\frac{  \lb K_2| 2\!\!\!/\ \mu\!\!\!/\
 1 \!\!\!/\ |K_1\rb }{   \lb 21\rb \lb 12 \rb}
 \ee As $\mu \!\!\!/\ $ anti-commutes with $1 \!\!\!/\ , 2 \!\!\!/\ $, those combine to give $(1+2)^2-1 \!\!\!/ 2 \!\!\!/\ $. The last term gives vanishing contribution (using e.g. $   [P| 2 \!\!\!/|K_1 \rb =0$ by momentum conservation), therefore
\be \A_4(K_1^-,K_2^-,1^+,2^+) =   \frac{[12]}{     \lb 12 \rb} \frac{  \lb K_2|   \mu\!\!\!/|K_1\rb }{(k_2+p_1)^2-\mu^2}
   \ee
 This   matches the result quoted in \cite{massive}. Similarly for the last  helicity configuration one obtains
\be \A_4(K_1^+,K_2^+,1^+,2^+) =   \frac{[12]}{     \lb 12 \rb}   \frac{[ K_2|   \mu\!\!\!/\
   |K_1]}{(k_2+p_1)^2-\mu^2}\ee

\subsection{Type-B Amplitudes}
These amplitudes have  two massive scalars (of opposite helicities)
with momenta $L_1,L_2$, and two massless fermions of opposite
helicities, and momenta $k_1= \lambda_1 \tilde{\lambda_1}$ and
$k_2=\lambda_2 \tilde{\lambda_2}$. As we are interested always in
adjacent massive legs, the only non-vanishing helicity preserving configurations  have
cyclic ordering of momenta $(k_1^+,k_2^-,L_1^-,L_2^+)$ or
$(k_1^+,k_2^-,L_1^+,L_2^-)$. The helicity violating  configuration is
$(k_1^+,k_2^+,L_1^+,L_2^+)$.

For all these configurations we choose the two marked
momenta to be $(i,j)=(k_1,L_1)$. In this case there are two
possible grouping of momenta, or two possible diagrams in the
recursion relation. In one of them the intermediate momentum is
$P=k_1+L_2$, and the intermediate state is a D-dimensional fermion.
 In the other diagram the intermediate momentum is $q=k_1+k_2$, and the intermediate state
 is a gluon which can be of positive or negative helicity. The first set of diagrams  can lead to helicity flipping via the D-dimensional internal fermion,
whereas the diagrams with an internal gluon only lead to helicity conserving amplitudes.

 The first amplitude $\A_4(k_1^+,k_2^-,L_1^-,L_2^+)$ can be written as
 a
sum of three terms \bea \A_4(k_1^+,k_2^-,L_1^-,L_2^+)=
-\frac{[\hat{k}_1 -\hat{P}]\lb \hat{P} k_2\rb}{P^2} - \frac{\lb k_2
q_1 \rb [\hat{k_1} -\hat{q}]}{\lb q_1 -\hat{q} \rb }\frac{1}{Q^2}
\frac{\lb \hat{q} | l_2| q_2]}{[q_2 \hat{q}]} + \nonumber \\
+\frac{\lb k_2 \hat{q} \rb [q_3 \hat{k}_1 ]}{[q_3
\hat{q}]}\frac{1}{Q^2} \frac{\lb q_4 | l_2 | \hat{q} ]}{\lb q_4
\hat{q} \rb} \nonumber \eea The last term vanishes if we choose
$q_3= k_1$ (note that $\lb k_1 \hat{q} \rb \neq 0$).  For the middle
term we choose $q_1=q_2 = k_1$, the amplitude is then \be
~~~~~~~~~~~~~~~~~~~~~~~~~~~~\frac{1}{q^2} \left\{[k_1| l_2| k_2\rb +
\frac{[\hat{k}_1 | \hat{q} \!\!\!/ l_2 \!\!\!/ k_1 \!\!\!/ | k_2
\rb}{\lb k_1 \hat{q} \rb [k_1 \hat{q} ]} \right\} \ee using
elementary consideration the numerator can be simplified
\be~~~~~~~~[\hat{k}_1 | \hat{q} \!\!\!/ l_2 \!\!\!/ k_1 \!\!\!/ |
k_2 \rb= -2(k_1 \cdot k_2)[k_1|l_2 \!\!\!/|k_2\rb = -2(k_1 \cdot
\hat{q})[k_1|l_2 \!\!\!/|k_2\rb \ee therefore \be
\A_4(k_1^+,k_2^-,L_1^-,L_2^+)= [k_1| l_2\!\!\!/| k_2\rb
\left\{\frac{1}{(k_1+l_2)^2 - \mu^2}+\frac{1}{(k_1+k_2)^2} \right\}
\ee which can be easily checked to be the result obtained from
Feynman diagrams.

Using similar reasoning, the   the other helicity conserving
amplitude  evaluates to be \be \A_4(k_1^+,k_2^-,L_1^+,L_2^-)=
\frac{[k_1|l_2\!\!\!/|k_2\rb}{(k_1+k_2)^2} \ee Note the absence of
singularity as $k_1+L_2$ becomes on -shell, as there is no
appropriate helicity assignment for the would-be on-shell
intermediate state. This result can be easily checked to be the
result of a sum of two Feynman diagrams.

As before, the intermediate state is effectively four dimensional for helicity preserving external states. We now evaluate the helicity flipping amplitude, for them the only possible intermediate state is fermionic and its   propagator is effectively $\mu \!\!\!/$, which simplifies the calculation. The primitive vertex coupling scalars to fermions is unmodified, giving:
\be\A_4(k_1^+,k_2^+,L_1^+,L_2^+)=
\frac{[\hat{k}_1 -\hat{P}][ \hat{P} k_2]}{P^2} =\frac{-[k_1|\mu\!\!\!/|k_2]}{(k_1+l_2)^2-\mu^2}\ee

However, for the purpose those type B amplitudes where the
``helicity" of the scalars is flipped are not used as ingredient in
the one-loop calculation we are ultimately interested in. We
therefore only consider the case where the two massive scalars have
opposite ``helicity". This is useful because often this eliminates
the fermion helicity-flipping terms as well.

\subsection{Type-C Amplitudes}
There are a few helicity configurations relevant here. We will
sketch the calculation and give the results. In all cases we take
the fermionic momenta to be $k_1,K_2$, the scalar momenta $L$  and
the gluon momentum $p$.

For $\A_4(k_1^+,K_2^+, L^+,p^+)$ we choose the marked momenta to be
$(i,j)=(p^+,k_1^+)$. There is one possible grouping of momenta for
which the intermediate state is a D-dimensional scalar of momentum
$Q= p+L$. The recursion relation reads \be
 \A_4(k_1^+,K_2^+, L^+,p^+)= -\frac{\lb q|L|\hat{p} ]}{\lb q
 \hat{p}\rb} \, \frac{1}{Q^2} \, [\hat{k}_1 K_2]
 \ee
 choosing $q=\hat{k}_1$ gives \be ~~~~~~~~~~~~~~~~~~~~~~~~~~~~\frac{\lb k_1|l\!\!\!/ |p] [K_2 \,\hat{k_1}]}{\lb k_1 p \rb \, Q^2}  \ee where we already projected onto the appropriate component of $L\!\!\!/$,namely $l\!\!\!/$.

The last step involved calculating $[K_2 \hat{k_1}]$ which will be
used repeatedly below. Using the result \be
~~~~~~~~~~~~~~~~~~~~~~~~~~~~[K_2 \hat{k_1}]= \frac{[p|\,l
\!\!\!/k_2\!\!\!/+\mu^2|K_2]}{\lb k_1|l\!\!\!/|p]} \ee one finally
gets

\be \A_4(k_1^+,K_2^+, L^+,p^+)=\frac{[p|\,l \!\!\!/k_2\!\!\!/+\mu^2|K_2] }{\lb k_1 p \rb \, \left[(p+l)^2-\mu^2 \right]} \ee

 For the amplitude $\A_4(k_1^+,K_2^+,L^+, p^-)$ we choose the marked
 momenta $(i,j)= (k_1^+,p^-)$, resulting in similar calculation (where
 the helicity of  the external gluon is flipped). The result
 is
 \be \A_4(k_1^+,K_2^+,L^+, p^-)= \frac{[k_1 K_2]\lb
 p|l\!\!\!/|k_1]}{[k_1 p]\left[ (p+l)^2-\mu^2 \right] }\ee

Note that for these two amplitudes, where the fermions are adjacent, the intermediate state is massive scalar and consequently there is no helicity-flipping intermediate state.

 For the amplitude $\A_4(k_1^+, p^+,K_2^+,L^+)$ we choose
 $(i,j)=(p^+,k_1^+)$, therefore the intermediate state is of momentum $Q=p+K_2$, and \be \A_4(k_1^+, p^+,K_2^+,L^+)=
 -\frac{[K_2 \hat{p}] \lb q -\hat{Q} \rb}{\lb q \hat{p} \rb}
 \,\frac{1}{Q^2} \, [\hat{Q} \hat{k}_1] \ee choosing $q=k_1$ gives
 after some elementary algebra \be \A_4(k_1^+, p^+,K_2^+,L^+)=
 \frac{\mu^2}{\lb k_1 p\rb \lb p K_2\rb} \ee

% FIX UP THIS ONE.

 As the intermediate state in this amplitude is D-dimensional fermion, there is a similar amplitude with opposite helicity for one of the fermions, namely
\be \A_4(k_1^+, p^+,K_2^-,L^+)=
  \frac{\lb K_2 q\rb   [\hat{p} \, \hat{Q} ]}{\lb  \hat{p} \,q\rb}
 \,\frac{1}{P^2} \, [\hat{P} \hat{k}_1] = \frac{\lb K_2 k_1\rb [p|\mu\!\!\!/|\hat{k_1}] }{\lb p \,k_1 \rb \, Q^2}\ee
after calculating  $[p|\mu\!\!\!/|\hat{k_1}]$ we end up with the amplitude
\be \A_4(k_1^+, p^+,K_2^-,L^+)=\frac{\lb K_2 k_1\rb [p|\mu\!\!\!/|k_1] }{\lb p \,k_1 \rb \, \left[(p+k_2)^2-\mu^2 \right]}\ee

 Finally for the amplitude $\A_4(k_1^+, p^-,K_2^+,L^+)$ we choose $(i,j)=(k_1^+,
 p^-)$, giving \be \A_4(k_1^+, p^-,K_2^+,L^+)= -[\hat{k}_1 -\hat{P}
 ] \frac{1}{P^2} \frac{\lb \hat{P} \hat{p} \rb [q K_2]}{[q \hat{p}]}
 \ee with $P=K_2+p$, choosing $q=k_1$ gives \be \A_4(k_1^+, p^-,K_2^+,L^+)=
 -\frac{[k_1 K_2] }{[k_1 p]} \frac {\lb p | \,l \!\!\!/|k_1]}{(k_1+l)^2-\mu^2}
\ee

Once again, since the intermediate state has helicity flipping part, we could get the additional amplitude
\be \A_4(k_1^+, p^-,K_2^-,L^+)= [\hat{k}_1 -\hat{P}
 ] \frac{1}{P^2} \frac{\lb K_2 \hat{p} \rb [q \hat{P}]}{[q \hat{p}]}
 \ee  However, choosing again $q=k_1$, and using  $[k_1|\mu\!\!\!/|k_1]=0$, this amplitude vanishes.
%gives
%\be \A_4(k_1^+, p^-,K_2^-,L^+)=\frac{[k_1|\mu\!\!\!/|k_1]}{[k_1 p][p K_2]} \ee

\section{Five Point Amplitudes}
We list below the results of the calculation of the relevant five
point amplitudes, and the checks they satisfy. In all cases the
calculation follows the lines of the corresponding four point
amplitude calculation, and we omit the details for brevity.
%In addition, different choices gives different representations of the
%same amplitudes, we always quote the most compact representation we
%found.

\subsection{Type-A Amplitudes}
These amplitudes include two adjacent D-dimensional fermions of
momenta   $K_1,K_2$,and three adjacent gluons of
momenta $p_1,p_2,p_3$.   The results are  \bea \A_5(K_1^-,K_2^-,
p_1^+,p_2^+,p_3^+) = \frac{\lb K_2|\mu \!\!\!/|K_1\rb\, \lb K_2
|(k_1\!\!\!/ + k_2 \!\!\!/)|3] }{\lb K_2 1 \rb \lb 1 2 \rb \lb 2 3
\rb \lb 3 K_1\rb [K_1 3] } \nonumber \eea There is no singularity as
$K_1+K_2$ becomes null, since no helicity assignment exists for an
intermediate state.   The result is symmetric
when exchanging $(1,3)$ and $(K_1,K_2)$, therefore there are only
two collinear limits to check. We perform these checks in the
appendix as a demonstration.

Note also that the amplitude vanishes in four
dimensions (setting $\mu=0$) as it should, and therefore only helicity-flipping parts exist. The other possible helicity assignment for fermions gives
\bea \A_5(K_1^+,K_2^+,
p_1^+,p_2^+,p_3^+) = \frac{[ K_1|(p_3\!\!\!/+k_1\!\!\!/)\mu \!\!\!/|2\rb\, \lb K_2
|(k_1\!\!\!/ + k_2 \!\!\!/)|3] }{\lb K_2 1 \rb \lb 1 2 \rb \lb 2 3
\rb \lb 3 K_1\rb [K_1 3] } \nonumber \eea

The rest of the type A amplitudes are given as sum over two diagrams, one with internal gluon and one with internal D-dimensional fermion

\bea\A_5(K_1^+,K_2^-, p_1^+,p_2^-,p_3^+) = -\frac{\lb K_2 2 \rb ^3
\, \lb K_1 2\rb [3 K_1]}{\lb K_2 1 \rb \lb 12 \rb \lb 23 \rb \lb 3
K_1 \rb \,[3|(k_1\!\!\!/+k_2\!\!\!/)| K_2 \rb} \nonumber \eea

%\nonumber \\+ \frac{\lb K_2 2 \rb}{\lb 21\rb \lb K_1 3\rb} \frac{[1|\mu\!\!\!/|3] \lb 2|\mu\!\!\!/|2\rb}
%{(K_2+p_1)^2 (p_2+p_3)^2}

\bea\A_5(K_1^-,K_2^-, p_1^+,p_2^-,p_3^+) = -
\frac{[13]^4\,\lb K_1|\mu\!\!\!/|K_2\rb}{[12][23][K_1 K_2] \lb K_2 K_1
\rb \,[3|(k_1 \!\!\!/+k_2\!\!\!/)|K_2 \rb}-\nonumber\\
\nonumber -\frac{\lb K_2 2 \rb \lb 2 K_1 \rb\,[3|k_1\!\!\!/|2\rb [1|\mu\!\!\!/|3] }{\lb 21\rb (K_2+p_1)^2 \, (p_2+p_3)^2\, (K_1+p_3)^2} \eea
%\left( + [1|k_2\!\!\!/|2\rb [3|\mu\!\!\!/|3] \right)
\bea\A_5(K_1^+,K_2^+, p_1^+,p_2^-,p_3^+) =  -
\frac{[13]^4\,[ K_1|\mu\!\!\!/|K_2]}{[12][23][K_1 K_2] \lb K_2 K_1
\rb \,[3|(k_1 \!\!\!/+k_2\!\!\!/)|K_2 \rb}\nonumber\eea
%\nonumber\\+ \frac{2 [1 K_2][K_1 3] [3|k_1 \!\!\!/|2 \rb \lb 2 |\mu  \!\!\!/| 2 \rb}{\lb 21 \rb (K_2+p_1)^2 \, (p_2+p_3)^2\, (K_1+p_3)^2}

\bea \A_5(K_1^+,K_2^-,
p_1^+,p_2^+,p_3^-) = \frac{\lb K_2 3 \rb ^2\, [K_1| (k_1
 \!\!\!/+k_2
\!\!\!/)| 3\rb}{\lb 12\rb \lb 23 \rb \lb K_1
K_2\rb \, [K_1| (k_1 \!\!\!/+k_2\!\!\!/)|1\rb} +\nonumber \\
+ \frac{\mu^2 [12] \lb K_2 3 \rb [2 K_1]}{\lb 12 \rb [23] \, (K_2 +p_1)^2 \, (K_1+p_3)^2} \nonumber \eea

\bea A_5(K_1^+, K_2^+,p_1^+,p_2^+,p_3^-)= \frac{[12][2 K_1]^2\,[K_1| \mu\!\!\!/|K_2 ]}{\lb K_2 1 \rb [K_2 1]
[23][K_1 3]\,[K_1| (k_2 \!\!\!/+ k_1 \!\!\!/ )|1 \rb }\nonumber\eea

 %\frac{[12] \lb K_2 K_1 \rb [K_1|
 %(k_1 \!\!\!/+k_2\!\!\!/)| 1\rb \, [ K_2|\, \mu\!\!\!/|K_1 ]}{\lb  2 3 \rb \lb 1 K_2 \rb \lb 1 3 \rb \, (K_1 +p_3)^2 \,(p_1+p_2)^2 }
%\nonumber \\
%+PROBLEMS WITH COLLINEAR LIMIT ON FIRST TERM.

\bea \A_5(K_1^+,K_2^-, p_1^+,p_2^-,p_3^-) = \frac{\lb 1 K_2 \rb ^2
\, [K_1|(k_1\!\!\!/ + k_2\!\!\!/)|1\rb\,[K_2|(k_1\!\!\!/ +
k_2\!\!\!/)|1\rb }{\lb K_2 1 \rb \lb 12 \rb \lb 23 \rb \lb K_1 K_2
\rb [K_2 K_1]\,[K_2|(k_1\!\!\!/ + k_2\!\!\!/)|3\rb }
\nonumber\eea

\bea \A_5(K_1^+,K_2^+, p_1^+,p_2^-,p_3^-)=
 \frac{[23] \, [K_2 2]^2 \lb K_2 1 \rb\,[K_1 | \mu \!\!\!/|K_2]}{
 (p_3+K_1)^2 \, (K_2+p_1)^2 \,[K_2| (k_1\!\!\!/+ k_2 \!\!\!/)|3 \rb}\nonumber\eea
%\nonumber \\  + \frac{[K_2 K_1] \lb K_1 1 \rb \lb 1| \mu |1\!\!\!/\rb }
%{\lb 21 \rb \lb 31 \rb \lb 32 \rb \lb K_2 K_1 \rb [1 K_2 ] }
%SAME PROBLEM WITH COLLINEAR LIMITS, SECOND LINE.

\subsection{Type-B Amplitudes}
These amplitudes have  two massive scalars (of opposite helicities)
with momenta $L_1,L_2$, and two massless fermions of  momenta $k_1,k_2$, and an additional gluon of
momentum $p$. The helicity flipping part of the fermion propagator are less relevant in this set of calculations, since fermion helicities are correlated with that of the external scalars.  The results are

\bea \A_5(p^+,L_1^-,L_2^+,k_1^+,k_2^-) = \frac{\lb
k_2|l_1\!\!\!/|p]\, [k_1| l_2\!\!\!/|k_2\rb }{\lb k_2 p \rb \,\lb
p|l_1\!\!\!/|p]}\left( \frac{\lb k_2 | l_1 \!\!\!/| p] }{[k_1|(l_2
\!\!\!/ l_1 \!\!\!/+\mu^2)|p] \,\lb k_1 k_2 \rb}
+\frac{1}{(k_1+L_2)^2} \right)\nonumber \\+ \frac{\mu^2 [p
k_1]^3}{[k_1|(l_2\!\!\!/ l_1\!\!\!/ +\mu^2)|p]\, [k_2 k_1]\,
(L_1+L_2)^2 } \nonumber \eea

 \bea \A_5(p^-,L_1^-,L_2^+,k_1^+,k_2^-)
=  \left( \frac{\lb p |l_2 \!\!\!/|k_2]}{\lb p|
(l_1\!\!\!/l_2\!\!\!/+\mu^2)|k_1\rb \, [k_1 k_2]}+
\frac{1}{(k_2+L_2)^2} \right) \times \,\,\,\,\,\,\,\,\,\,\,\,\,\,\,\,\nonumber \\ \nonumber\\
  \times \left(\frac{[k_1|l_1\!\!\!/|p\rb \,
[k_1|l_2\!\!\!/|k_2 \rb - \mu^2 [12]\lb2p\rb}{\lb p|l_1\!\!\!/|p] \,
[p2]}\right) +\frac{\mu^2 \, \lb p k_2 \rb^2 \, \lb p k_1 \rb}{\lb p
|(l_1\!\!\!/l_2\!\!\!/+\mu^2)|k_2\rb\,\lb k_1 k_2 \rb \,
(L_1+L_2)^2} \nonumber\eea

\bea \A_5(p^+,L_1^-,L_2^+,k_1^-,k_2^+) = \frac{\lb
k_2|l_1\!\!\!/|p]^2 \, \lb k_1 |l_2\!\!\!/|k_2] }{\lb k_1 k_2 \rb
\lb k_2 p \rb \, \lb p|l_1 \!\!\!/|p] \,
[k_1|(l_2\!\!\!/l_1\!\!\!/+\mu^2)|p]} \nonumber
\\ + \frac{\mu^2\,[p k_2]^2 \, [p k_1]}
{[k_1 k_2] \, [k_1|(l_2\!\!\!/l_1\!\!\!/+\mu^2) |p]\, (L_1+L_2)^2 }
\nonumber \eea

\bea   \A_5(p^-,L_1^-,L_2^+,k_1^-,k_2^+)= -\frac{\lb
p|l_1\!\!\!/|k_2]^2 \, \lb k_1 |l_2\!\!\!/|k_2] }{ [k_1 k_2][k_2 p]
\lb p|l_1 \!\!\!/|p] \, \lb p|(l_1\!\!\!/l_2\!\!\!/+\mu^2)|k_1\rb}
\nonumber
\\ + \frac{\mu^2\,  \lb p k_1\rb^3}
{\lb k_1 k_2 \rb  \, \lb p|(l_1\!\!\!/l_2\!\!\!/+\mu^2)|k_1\rb \,
(L_1+L_2)^2 } \nonumber \eea

\bea \A_5(L_1^+,p^+,L_2^-,k_1^-,k_2^+)= \frac{\lb k_2|l_1\!\!\!/ l_2
\!\!\!/|k_1 \rb}{\lb k_1 p \rb \lb p k_2 \rb}\left(\frac{\lb k_1 k_2
\rb}{\lb k_2 | l_1\!\!\!/ l_2 \!\!\!/|k_1 \rb}+\frac{1}{(L_1+L_2)^2}
\right)\nonumber\\ +\frac{\mu^2 \, [p|l_1\!\!\!/|k_2 \rb^2}{\lb
k_2|l_1\!\!\!/ l_2 \!\!\!/|k_1
\rb\,(L_1+L_2)^2\,(L_1+k_2)^2}\nonumber \eea

\bea \A_5(L_1^-,p^+,L_2^+,k_1^-,k_2^+)= \frac{\lb k_2|l_2\!\!\!/ l_1
\!\!\!/|k_1 \rb}{\lb k_1 p \rb \lb p k_2 \rb\,(L_1+L_2)^2} \nonumber
\eea

We also find that all amplitude where fermion helicity is flipped are vanishing.

\subsection{Type-C Amplitudes}
These consist of massless fermion of momentum $k_1$, massive fermion of momentum $K_2$,
  massive scalar of momentum $L$  and two massless gluons of
momenta $p_1,p_2$. The results are: \bea \A_5(p_1^+,p_2^+,k_1^+,L^+,K_2^+)= -\frac{\mu^2}{\lb K_2 1\rb \lb 12 \rb \lb 2 k_1\rb} \nonumber\eea

\bea \A_5(p_1^+,p_2^+,k_1^+,L^+,K_2^-)= \frac{\lb
K_2|k_1\sla\mu\sla(l\sla+k\sla_2)|K_2\rb}{\lb K_2 1\rb \lb 12 \rb
\lb 2 k_1\rb(k_1+L)^2} \nonumber\eea

\bea \A_5(p_1^+,p_2^-,k_1^+,L^+,K_2^+)= - \frac{\lb 2 K_2 \rb [k_1|l
\!\!\!/|2 \rb ^2} {\lb K_2 1 \rb \lb 12\rb
\,[k_1|(l\!\!\!/+k_2\!\!\!/)|K_2 \rb \, (k_1+L)^2} \nonumber \\-
\frac{\mu^2 [k_1 1]^3}{[12][2 k_1]\,[k_1|(l\!\!\!/+k_2\!\!\!/)|K_2
\rb \, (k_1+L)^2} \nonumber \eea

\bea \A_5(p_1^+,p_2^-,k_1^+,L^+,K_2^-)= \frac{\lb K_2
2\rb^2[1|\mu\sla|k_1]\lb 2|l\sla|k_1]}{\lb K_2 1\rb\lb 1
2\rb[k_1|l\sla| K_2\rb[K_2 1](k_1+L)^2} \nonumber \\ +\frac{\lb
K_2|(l\sla+k_2\sla)|1][1|\mu\sla|k_1][1 k_1]^2}{[K_2 1][1 2][2
k_1][k_1|(l\sla+k_2\sla)|K_2\rb(K_2+L)^2} \nonumber\eea

\bea \A_5(p_1^-,p_2^+,k_1^+,L^+,K_2^+)=  \frac{\lb 1 k_1 \rb \lb 1
|l \!\!\!/ +k_2\!\!\!/|K_2] ^2} {\lb 12 \rb \lb 2 k_1\rb \,\lb
k_1|(l\!\!\!/+k_2\!\!\!/)|K_2 ] \, (K_2+L)^2} \nonumber \\-
\frac{\mu^2 [K_2 2]^3}{[12][K_2 1]\,[K_2|(l\!\!\!/+k_2\!\!\!/)|k_1
\rb \, (k_1+L)^2} \nonumber \eea

\bea \A_5(p_1^-,p_2^+,k_1^+,L^+,K_2^-)=  \frac{\lb
K_1|l\sla|2][2|\mu\sla|k_1][K_2 2]} {[K_2|(l\sla+k_2\sla)|k_1\rb
[K_2 1][12](k_1+L)^2} \nonumber\eea

\bea \A_5(p_1^-,p_2^-,k_1^+,L^+,K_2^+)= \frac{[K_2 k_1] \, [k_1|
l\!\!\!/ \mu \!\!\!/ +(k_1+L)^2|K_2]}{[12][K_2 1][2
k_1]\,(k_1+L)^2}\nonumber \eea

%DIFFERENCES BETWEEN l AND L??

\bea \A_5(p_1^+,k_1^+,p_2^+,L^+,K_2^+)= -\frac{\mu^2\lb
k_1|l\sla|2]}{\lb K_2 1\rb \lb 1k_1 \rb \lb k_1 2\rb\lb 2|l\sla|2]}
\nonumber\eea

\bea \A_5(p_1^+,k_1^+,p_2^+,L^+,K_2^-)= -\frac{\lb K_2
k_1\rb[1|\mu\sla(k_2\sla l\sla+\mu^2)|2]}{\lb K_2 1\rb \lb 1k_1 \rb
\lb k_1 2\rb\lb 2|l\sla|2][1 K_2]} \nonumber\eea

\bea \A_5(p_1^+,k_1^+,p_2^-,L^+,K_2^+)=  \frac{\lb 2 K_2 \rb
\,[K_2|l \!\!\!/|2 \rb[K_2|l \!\!\!/+k_2\!\!\!/ |2 \rb } {\lb  1 k_1
\rb \  \, \lb 2|(l\!\!\!/ k_2\!\!\!/+\mu^2)|1 \rb \, (K_2+L)^2}
\nonumber \\+ \frac{\mu^2  \lb 2| l\!\!\!/|k_1]^2}{[k_1 2]\lb K_2
1\rb\,\lb 2|l\!\!\!/|2]\,\lb 2|(l\!\!\!/ K_2\!\!\!/+\mu^2)|1 \rb }
\nonumber \eea

\bea \A_5(p_1^+,k_1^+,p_2^-,L^+,K_2^-)= \frac{\lb K_2|((k_2\sla
+1\sla)l\sla+\mu^2)|2\rb[1|\mu\sla|k_1]\lb 2|l\sla|k_1]}{\lb
1|(k_2\sla l\sla+\mu^2)|2\rb\lb 1 K_2\rb[k_2 1][k_1 2]\lb
2|l\sla|2]}\nonumber\eea

\bea \A_5(p_1^-,k_1^+,p_2^+,L^+,K_2^+)=
\frac{[2|l\sla(k_1\sla+2\sla)|K_2]^2}{[K_2 1]\lb k_1
2\rb\lb2|l\sla|2][1|(k_2\sla l\sla+\mu^2)|2]}\nonumber
\\+\frac{[k_1 2]^2[2|(l\sla k_2\sla+\mu^2)|K_2]}{[1 k_1][2|(l\sla
k_2\sla+\mu^2)|1](K_2+L)^2}\nonumber\eea

\bea \A_5(p_1^-,k_1^+,p_2^-,L^+,K_2^+)= \frac{[K_2
k_1]^2\lb2|l\sla|k_1]}{[K_2 1][1 k_1][k_1
2]\lb2|l\sla|2]}\nonumber\eea

\bea \A_5(k_1^+,p_1^+,p_2^+,L^+,K_2^+)= \frac{[2|(l\sla
k_2\sla+\mu^2)|k_1]\lb k_1|k_2\sla|K_2]}{\lb k_1
1\rb\lb12\rb\lb2|l\sla|2][(K_2+k_1)^2-\mu^2]}\nonumber\eea

\bea \A_5(k_1^+,p_1^+,p_2^+,L_+,K_2^-)= \frac{\lb1|l\sla|2]\lb
K_2k_1\rb[2|(l\sla k_2\sla+\mu^2)|k_1]}{\lb
k_11\rb\lb12\rb\lb2|l\sla|2][2|(l\sla k_2\sla+\mu^2)|K_2]\lb
K_21\rb}\nonumber\eea

\bea \A_5(k_1^+,p_1^+,p_2^-,L^+,K_2^+)= \frac{\lb
k_12\rb[K_2|(l\sla+k_2\sla)|2\rb\lb2|l\sla|K_2]\lb K_22\rb}{\lb
k_11\rb\lb12\rb\lb2|(l\sla
k_2\sla+\mu^2)|k_1\rb(K_2+L)^2}\nonumber\\ +\frac{\lb
k_1|k_2\sla|K_2][k_11]\lb2|l\sla|1]^2}{[12][2|l\sla|2\rb\lb
k_1|(k_2\sla l\sla+\mu^2)|2\rb(k_1+K_2)^2}\nonumber\eea

\bea \A_5(k_1^+,p_1^+,p_2^-,L^+,K_2^-)= \frac{\lb
K_2k_1\rb[1|\mu\sla|k_1]\lb2|l\sla|1]^3}{[K_21][12][2|l\sla|2\rb\lb2|l\sla(1\sla+2\sla)|K_2\rb\lb2|(l\sla
k_2\sla+\mu^2)|k_1\rb} \nonumber\eea

\bea \A_5(k_1^+,p_1^-,p_2^+,L^+,K_2^+)=
\frac{\lb1|l\sla|2]^2[k_1K_2]^2\lb1K_2\rb}{\lb12\rb\lb2|l\sla|2][k_1|(k_2\sla
l\sla+\mu^2)|2][(k_1+K_2)^2-\mu^2]}\nonumber \\ +\frac{[2|(l\sla
k_2\sla+\mu^2)|K_2][2k_1]^3}{[k_11][12][2|(l\sla
k_2\sla+\mu^2)|k_1](K_2+L)^2}\nonumber\eea

\bea \A_5(k_1^+,p_1^-,p_2^+,L^+,K_2^-)= \frac{\lb
K_2|(k_2\sla+l\sla)|2][2|\mu\sla|k_1][2k_1]^2}{[k_11][12][2|l\sla
k_2\sla+\mu^2|k_1](K_2+L)^2} \nonumber\eea

\bea \A_5(k_1^+,p_1^+,p_2^+,L^+,K_2^+)=
\frac{[k_1K_2]^2\lb2|l\sla(1\sla+2\sla)|K_2\rb}{[k_11][12][2|l\sla|2\rb[(k_1+K_2)^2-\mu^2]}\nonumber\eea

To summarize, we have applied the BCF recursion relations to
amplitudes which involve D-dimensional fermions and scalars.  These
D-dimensional particles behave in most respects as their massive
counterparts in four dimensions.  We have used this formalism to
obtain four and five point amplitudes where two of the particles
have been continued away from four dimensions and the remaining
particles are (on-shell) gluons and massless fermions.  Our results
posses expected factorization properties and have passed other
consistency checks, as outlined in section 2.6.  The tree-level
amplitudes we have computed here are the building blocks (to be
assembled using generalized unitarity in D dimensions) of the
rational (nonsupersymmetric) terms of the one-loop amplitude of two
(massless) fermions and up to three gluons.   We leave this task for
future work \cite{progress}.

\section*{Appendix}

Let us discuss the factorization limits of the amplitude \bea
\A_5(K_1^-,K_2^-, p_1^+,p_2^+,p_3^+) = \frac{\lb K_2|\mu
\!\!\!/|K_1\rb\, \lb K_2 |(k_1\!\!\!/ + k_2 \!\!\!/)|3] }{\lb K_2 1
\rb \lb 1 2 \rb \lb 2 3 \rb \lb 3 K_1\rb [K_1 3] } \nonumber \eea

As mentioned in the text, there is no singularity in the channel
where $K_1+K_2$ becomes on shell. In addition there is a symmetry of
exchanging $(1,3)$ and $(K_1,K_2)$. This leaves two channels to
check, when $p_1+p_2$ becomes null, or when $p_3+K_1$ becomes null
in D-dimensions (so it approaches the mass shell condition
$(p_3+k_1)^2=\mu^2$ in four dimensions).

For the first limit we denote $p=p_1+p_2$, the amplitude as $p$
becomes light-like  factorizes to \bea
\A_5(K_1^-,K_2^-,p_1^+,p_2^+,p_3^+) \rightarrow
\A_3(p_1^+,p_2^+,-p^-) \, \frac{1}{p^2} \,
\A_4(p^+,p_3^+,K_1^-,K_2^-) \nonumber \eea The four point amplitude
is of type A, so we get: \bea \A_5(K_1^+,K_2^-, p_1^+,p_2^+,p_3^+)
\rightarrow \frac{[12]^3}{[1p][p2]}\,\frac{1}{\lb12\rb
[21]}\,\frac{[p 3]}{\lb p 3 \rb}\frac{\lb K_1|\mu
\!\!\!/|K_2\rb}{(K_1+p_3)^2} \nonumber \eea To get to the right form
we multiply both numerator and denominator by $\lb p K_2 \rb$, and
use momentum conservation to eliminate $p$, for example \bea [1p]
\lb p 3\rb= [1|p|3\rb = [1|2|3 \rb= [12]\lb 23 \rb \nonumber \eea
this results in \bea \A_5(K_1^+,K_2^-, p_1^+,p_2^+,p_3^+)
\rightarrow \frac{[12]^3}{[12][12]}\,\frac{1}{\lb12\rb
[21]}\,\frac{\lb K_2 |(k_1\!\!\!/ + k_2\!\!\!/)|3] }{\lb K_2 1 \rb
\lb 2 3 \rb}\frac{\lb K_1|\mu \!\!\!/|K_2\rb}{(K_1+p_3)^2} \nonumber
\eea which coincides indeed with the factorization limit of the
exact expression calculated (generally there are additional terms in
the exact expression which are non-singular in the limit).

The second factorization limit is for the channel $Q=p_3+K_1$, in
that limit \bea \A_5(K_1,K_2, p_1^+,p_2^+,p_3^+) \rightarrow
\A_3(p_3^+,K_1, -Q) \, \frac{1}{Q^2} \, \A_4(Q,K_2,p_1^+,p_2^+)
\nonumber \eea where $Q$ is the momentum of a D-dimensional
intermediate fermion. However   physical amplitudes such as
$\A_5(K_1,K_2, p_1^+,p_2^+,p_3^+)$ are to be assembled from their
components such as  $\A_5(K_1^\pm,K_2^\pm, p_1^+,p_2^+,p_3^+)$ which
do not separately obey factorization constraints. We therefore did
not check these more complicated factorization limits involving
massive intermediate momentum.

\section*{Acknowledgements}
  We are
supported by National Science and Engineering Research Council of
 Canada.


\begin{thebibliography}{99}
%\cite{Witten:2003nn}
\bibitem{Witten}
  E.~Witten,
  ``Perturbative gauge theory as a string theory in twistor space,''
  Commun.\ Math.\ Phys.\  {\bf 252}, 189 (2004)
  [arXiv:hep-th/0312171].
  %%CITATION = HEP-TH 0312171;%%
  %\cite{Nair:1988bq}
\bibitem{Nair}
  V.~P.~Nair,
  ``A Current Algebra For Some Gauge Theory Amplitudes,''
Phys.\ Lett.\ B {\bf 214}, 215 (1988).
  %%CITATION = PHLTA,B214,215;%%
  %\cite{Parke:1986gb}
\bibitem{PT}
  S.~J.~Parke and T.~R.~Taylor,
  ``An Amplitude For N Gluon Scattering,''
  Phys.\ Rev.\ Lett.\  {\bf 56}, 2459 (1986).
  %%CITATION = PRLTA,56,2459;%%
  %\cite{Penrose:1972ia}
\bibitem{twistor}
  R.~Penrose and M.~A.~H.~MacCallum,
  ``Twistor Theory: An Approach To The Quantization Of Fields And Space-Time,''
  Phys.\ Rept.\  {\bf 6}, 241 (1972).
  %%CITATION = PRPLC,6,241;%%
%\cite{Cachazo:2004kj}
\bibitem{csw}
F.~Cachazo, P.~Svrcek and E.~Witten, ``MHV vertices and tree
amplitudes in gauge theory,'' JHEP {\bf 0409}, 006 (2004)
[arXiv:hep-th/0403047].
%%CITATION = HEP-TH 0403047;%%
%\cite{Georgiou:2004by}
\bibitem{tree}
C.~J.~Zhu, ``The googly amplitudes in gauge theory,'' JHEP {\bf
0404}, 032 (2004) [arXiv:hep-th/0403115];

G.~Georgiou and V.~V.~Khoze, ``Tree amplitudes in gauge theory as
scalar MHV diagrams,'' JHEP {\bf 0405}, 070 (2004)
[arXiv:hep-th/0404072];

G.~Georgiou, E.~W.~N.~Glover and V.~V.~Khoze, ``Non-MHV tree
amplitudes in gauge theory,'' JHEP {\bf 0407}, 048 (2004)
[arXiv:hep-th/0407027];

J.~B.~Wu and C.~J.~Zhu, ``MHV vertices and scattering amplitudes in
gauge theory,'' JHEP {\bf 0407}, 032 (2004) [arXiv:hep-th/0406085];

J.~B.~Wu and C.~J.~Zhu, ``MHV vertices and fermionic scattering
amplitudes in gauge theory with quarks and gluinos,'' JHEP {\bf
0409}, 063 (2004) [arXiv:hep-th/0406146];

D.~A.~Kosower, ``Next-to-maximal helicity violating amplitudes in
gauge theory,'' arXiv:hep-th/0406175;

X.~Su and J.~B.~Wu, ``Six-quark amplitudes from fermionic MHV
vertices,'' arXiv:hep-th/0409228;

%\cite{Birthwright:2005ak}
 T.~G.~Birthwright, E.~W.~N.~Glover, V.~V.~Khoze and P.~Marquard,
  ``Multi-gluon collinear limits from MHV diagrams,''
  JHEP {\bf 0505}, 013 (2005)
  [arXiv:hep-ph/0503063];
  %%CITATION = HEP-PH 0503063;%%

  %\cite{Birthwright:2005vi}
 T.~G.~Birthwright, E.~W.~N.~Glover, V.~V.~Khoze and P.~Marquard,
  ``Collinear limits in QCD from MHV rules,''
  JHEP {\bf 0507}, 068 (2005)
  [arXiv:hep-ph/0505219].
  %%CITATION = HEP-PH 0505219;%%




%\cite{Brandhuber:2004yw}
\bibitem{loop}
A.~Brandhuber, B.~Spence and G.~Travaglini, ``One-loop gauge theory
amplitudes in N = 4 super Yang-Mills from MHV vertices,''
arXiv:hep-th/0407214;

%\cite{Cachazo:2004by}
 F.~Cachazo, P.~Svrcek and E.~Witten,
  ``Gauge theory amplitudes in twistor space and holomorphic anomaly,''
  JHEP {\bf 0410}, 077 (2004)
  [arXiv:hep-th/0409245];
  %%CITATION = HEP-TH 0409245;%%

%\cite{Luo:2004ss}
M.~x.~Luo and C.~k.~Wen,
  ``One-loop maximal helicity violating amplitudes in N = 4 super Yang-Mills
  theories,''
  JHEP {\bf 0411}, 004 (2004)
  [arXiv:hep-th/0410045];
  %%CITATION = HEP-TH 0410045;%%


  I.~Bena, Z.~Bern, D.~A.~Kosower and R.~Roiban,
  ``Loops in twistor space,''
  Phys.\ Rev.\ D {\bf 71}, 106010 (2005)
  [arXiv:hep-th/0410054];
  %%CITATION = HEP-TH 0410054;%%


  M.~x.~Luo and C.~k.~Wen,
  ``Systematics of one-loop scattering amplitudes in N = 4 super Yang-Mills
  theories,''
  Phys.\ Lett.\ B {\bf 609}, 86 (2005)
  [arXiv:hep-th/0410118];
  %%CITATION = HEP-TH 0410118;%%

  %\cite{Quigley:2004pw}
 C.~Quigley and M.~Rozali,
  ``One-loop MHV amplitudes in supersymmetric gauge theories,''
  JHEP {\bf 0501}, 053 (2005)
  [arXiv:hep-th/0410278];
  %%CITATION = HEP-TH 0410278;%%

  %\cite{Bedford:2004py}
 J.~Bedford, A.~Brandhuber, B.~J.~Spence and G.~Travaglini,
  ``A twistor approach to one-loop amplitudes in N = 1 supersymmetric
  Yang-Mills theory,''
  Nucl.\ Phys.\ B {\bf 706}, 100 (2005)
  [arXiv:hep-th/0410280];
  %%CITATION = HEP-TH 0410280;%%

  %\cite{Bedford:2004nh}
J.~Bedford, A.~Brandhuber, B.~J.~Spence and G.~Travaglini,
  ``Non-supersymmetric loop amplitudes and MHV vertices,''
  Nucl.\ Phys.\ B {\bf 712}, 59 (2005)
  [arXiv:hep-th/0412108].
  %%CITATION = HEP-TH 0412108;%%



  %\cite{Dixon:2004za}
\bibitem{massive1}
  L.~J.~Dixon, E.~W.~N.~Glover and V.~V.~Khoze,
  ``MHV rules for Higgs plus multi-gluon amplitudes,''
  JHEP {\bf 0412}, 015 (2004)
  [arXiv:hep-th/0411092];
  %%CITATION = HEP-TH 0411092;%%

  %\cite{Bern:2004ba}
Z.~Bern, D.~Forde, D.~A.~Kosower and P.~Mastrolia,
  ``Twistor-inspired construction of electroweak vector boson currents,''
  Phys.\ Rev.\ D {\bf 72}, 025006 (2005)
  [arXiv:hep-ph/0412167];
  %%CITATION = HEP-PH 0412167;%%

  %\cite{Badger:2004ty}
  S.~D.~Badger, E.~W.~N.~Glover and V.~V.~Khoze,
  ``MHV rules for Higgs plus multi-parton amplitudes,''
  JHEP {\bf 0503}, 023 (2005)
  [arXiv:hep-th/0412275].
  %%CITATION = HEP-TH 0412275;%%

  %\cite{Bjerrum-Bohr:2005jr}
\bibitem{Bjerrum-Bohr:2005jr}
  N.~E.~J.~Bjerrum-Bohr, D.~C.~Dunbar, H.~Ita, W.~B.~Perkins and K.~Risager,
  ``MHV-vertices for gravity amplitudes,''
  arXiv:hep-th/0509016.
  %%CITATION = HEP-TH 0509016;%%


%\cite{Britto:2004ap}
\bibitem{BCF}
  R.~Britto, F.~Cachazo and B.~Feng,
  ``New recursion relations for tree amplitudes of gluons,''
  Nucl.\ Phys.\ B {\bf 715}, 499 (2005)
  [arXiv:hep-th/0412308].
  %%CITATION = HEP-TH 0412308;%%

  %\cite{Britto:2005fq}
\bibitem{BCFW}
  R.~Britto, F.~Cachazo, B.~Feng and E.~Witten,
  ``Direct proof of tree-level recursion relation in Yang-Mills theory,''
  Phys.\ Rev.\ Lett.\  {\bf 94}, 181602 (2005)
  [arXiv:hep-th/0501052].
  %%CITATION = HEP-TH 0501052;%%

  %\cite{Roiban:2004ix}
\bibitem{diss}
  R.~Roiban, M.~Spradlin and A.~Volovich,
  ``Dissolving N = 4 loop amplitudes into QCD tree amplitudes,''
  Phys.\ Rev.\ Lett.\  {\bf 94}, 102002 (2005)
  [arXiv:hep-th/0412265].
  %%CITATION = HEP-TH 0412265;%%


  %\cite{Luo:2005rx}
\bibitem{treebcf}
  M.~x.~Luo and C.~k.~Wen,
  ``Recursion relations for tree amplitudes in super gauge theories,''
  JHEP {\bf 0503}, 004 (2005)
  [arXiv:hep-th/0501121];
  %%CITATION = HEP-TH 0501121;%%

  %\cite{Luo:2005my}
 M.~x.~Luo and C.~k.~Wen,
  ``Compact formulas for all tree amplitudes of six partons,''
  Phys.\ Rev.\ D {\bf 71}, 091501 (2005)
  [arXiv:hep-th/0502009];
  %%CITATION = HEP-TH 0502009;%%

  %\cite{Britto:2005dg}
 R.~Britto, B.~Feng, R.~Roiban, M.~Spradlin and A.~Volovich,
  ``All split helicity tree-level gluon amplitudes,''
  Phys.\ Rev.\ D {\bf 71}, 105017 (2005)
  [arXiv:hep-th/0503198];
  %%CITATION = HEP-TH 0503198;%%








%\cite{Bedford:2005yy}
\bibitem{gravitybcf}
  J.~Bedford, A.~Brandhuber, B.~J.~Spence and G.~Travaglini,
  ``A recursion relation for gravity amplitudes,''
  Nucl.\ Phys.\ B {\bf 721}, 98 (2005)
  [arXiv:hep-th/0502146];
  %%CITATION = HEP-TH 0502146;%%

  %\cite{Cachazo:2005ca}
    F.~Cachazo and P.~Svrcek,
  ``Tree level recursion relations in general relativity,''
  arXiv:hep-th/0502160.
  %%CITATION = HEP-TH 0502160;%%




%\cite{Badger:2005zh}
\bibitem{massbcf}
  S.~D.~Badger, E.~W.~N.~Glover, V.~V.~Khoze and P.~Svrcek,
  ``Recursion relations for gauge theory amplitudes with massive particles,''
  JHEP {\bf 0507}, 025 (2005)
  [arXiv:hep-th/0504159];
  %%CITATION = HEP-TH 0504159;%%

  %\cite{Badger:2005jv}
\bibitem{massbcf2}
 S.~D.~Badger, E.~W.~N.~Glover and V.~V.~Khoze,
  ``Recursion relations for gauge theory amplitudes with massive vector bosons
  and fermions,''
  arXiv:hep-th/0507161;
  %%CITATION = HEP-TH 0507161;%%

  %\cite{Forde:2005ue}
\bibitem{massmore}
 D.~Forde and D.~A.~Kosower,
  ``All-multiplicity amplitudes with massive scalars,''
  arXiv:hep-th/0507292;
  %%CITATION = HEP-TH 0507292;%%

  %\cite{Rodrigo:2005eu}
  G.~Rodrigo,
  ``Multigluonic scattering amplitudes of heavy quarks,''
  arXiv:hep-ph/0508138.
  %%CITATION = HEP-PH 0508138;%%

%\cite{Bern:2005hs}
\bibitem{rational}
  Z.~Bern, L.~J.~Dixon and D.~A.~Kosower,
  ``On-shell recurrence relations for one-loop QCD amplitudes,''
  Phys.\ Rev.\ D {\bf 71}, 105013 (2005)
  [arXiv:hep-th/0501240];
  %%CITATION = HEP-TH 0501240;%%

  %\cite{Bern:2005ji}
 Z.~Bern, L.~J.~Dixon and D.~A.~Kosower,
``The last of the finite loop amplitudes in QCD,''
  arXiv:hep-ph/0505055;
  %%CITATION = HEP-PH 0505055;%%

  %\cite{Bern:2005hh}
 Z.~Bern, N.~E.~J.~Bjerrum-Bohr, D.~C.~Dunbar and H.~Ita,
  ``Recursive calculation of one-loop QCD integral coefficients,''
  arXiv:hep-ph/0507019.
  %%CITATION = HEP-PH 0507019;%%




%\cite{Risager:2005vk}
\bibitem{Risager:2005vk}
  K.~Risager,
  ``A direct proof of the CSW rules,''
  arXiv:hep-th/0508206.
  %%CITATION = HEP-TH 0508206;%%

  %\cite{Bern:1994zx}
\bibitem{Bern:1994zx}
  Z.~Bern, L.~J.~Dixon, D.~C.~Dunbar and D.~A.~Kosower,
  ``One loop n point gauge theory amplitudes, unitarity and collinear limits,''
  Nucl.\ Phys.\ B {\bf 425}, 217 (1994)
  [arXiv:hep-ph/9403226].
  %%CITATION = HEP-PH 9403226;%%

  %\cite{Dixon:1996wi}
\bibitem{Dixon:1996wi}
  L.~J.~Dixon,
  ``Calculating scattering amplitudes efficiently,''
  arXiv:hep-ph/9601359.
  %%CITATION = HEP-PH 9601359;%%

%\cite{Britto:2004nc}
\bibitem{gen}
  R.~Britto, F.~Cachazo and B.~Feng,
  ``Generalized unitarity and one-loop amplitudes in N = 4 super-Yang-Mills,''
  Nucl.\ Phys.\ B {\bf 725}, 275 (2005)
  [arXiv:hep-th/0412103];
  %%CITATION = HEP-TH 0412103;%%

  Z.~Bern, V.~Del Duca, L.~J.~Dixon and D.~A.~Kosower, ``All
Non-Maximally-Helicity-Violating One-Loop Seven-Gluon Amplitudes in
N=4 Super-Yang-Mills Theory,'' arXiv:hep-th/0410224;
%%CITATION = HEP-TH 0410224;%%


%\cite{Bern:2004bt}
 Z.~Bern, L.~J.~Dixon and D.~A.~Kosower,
  ``All next-to-maximally helicity-violating one-loop gluon amplitudes in N  =
  4 super-Yang-Mills theory,''
  Phys.\ Rev.\ D {\bf 72}, 045014 (2005)
  [arXiv:hep-th/0412210].
  %%CITATION = HEP-TH 0412210;%%


  %\cite{Bidder:2005ri}
 S.~J.~Bidder, N.~E.~J.~Bjerrum-Bohr, D.~C.~Dunbar and W.~B.~Perkins,
  ``One-loop gluon scattering amplitudes in theories with N < 4
  supersymmetries,''
  Phys.\ Lett.\ B {\bf 612}, 75 (2005)
  [arXiv:hep-th/0502028].
  %%CITATION = HEP-TH 0502028;%%


%\cite{Britto:2005ha}
 R.~Britto, E.~Buchbinder, F.~Cachazo and B.~Feng,
  ``One-loop amplitudes of gluons in SQCD,''
  arXiv:hep-ph/0503132.
  %%CITATION = HEP-PH 0503132;%%

%\cite{Brandhuber:2005jw}
\bibitem{pure}
  A.~Brandhuber, S.~McNamara, B.~J.~Spence and G.~Travaglini,
  ``Loop amplitudes in pure Yang-Mills from generalised unitarity,''
  arXiv:hep-th/0506068.
  %%CITATION = HEP-TH 0506068;%%

  %\cite{Bern:1994cg}
\bibitem{fusing}
Z.~Bern, L.~J.~Dixon, D.~C.~Dunbar and D.~A.~Kosower, ``Fusing gauge
theory tree amplitudes into loop amplitudes,'' Nucl.\ Phys.\ B {\bf
435}, 59 (1995) [arXiv:hep-ph/9409265].
%%CITATION = HEP-PH 9409265;%%

  %\cite{Bern:2005cq}
\bibitem{boot}
  Z.~Bern, L.~J.~Dixon and D.~A.~Kosower,
  ``Bootstrapping multi-parton loop amplitudes in QCD,''
  arXiv:hep-ph/0507005.
  %%CITATION = HEP-PH 0507005;%%

%\cite{Bern:1993mq}
\bibitem{five}
  Z.~Bern, L.~J.~Dixon and D.~A.~Kosower,
  ``One loop corrections to five gluon amplitudes,''
  Phys.\ Rev.\ Lett.\  {\bf 70}, 2677 (1993)
  [arXiv:hep-ph/9302280].
  %%CITATION = HEP-PH 9302280;%%

  %\cite{Bern:1994fz}
 Z.~Bern, L.~J.~Dixon and D.~A.~Kosower,
  ``One loop corrections to two quark three gluon amplitudes,''
  Nucl.\ Phys.\ B {\bf 437}, 259 (1995)
  [arXiv:hep-ph/9409393].
  %%CITATION = HEP-PH 9409393;%%

  %\cite{Kunszt:1994tq}
Z.~Kunszt, A.~Signer and Z.~Trocsanyi,
  ``One loop radiative corrections to the helicity amplitudes of QCD processes
  involving four quarks and one gluon,''
  Phys.\ Lett.\ B {\bf 336}, 529 (1994)
  [arXiv:hep-ph/9405386].
  %%CITATION = HEP-PH 9405386;%%

\bibitem{progress} Callum Quigley and Moshe Rozali, in progress.



%\cite{Mangano:1990by}
\bibitem{multi}
  M.~L.~Mangano and S.~J.~Parke,
  ``Multiparton Amplitudes In Gauge Theories,''
  Phys.\ Rept.\  {\bf 200}, 301 (1991).
  %%CITATION = PRPLC,200,301;%%



%\cite{Bern:1995db}
\bibitem{massive}
  Z.~Bern and A.~G.~Morgan,
  ``Massive Loop Amplitudes from Unitarity,''
  Nucl.\ Phys.\ B {\bf 467}, 479 (1996)
  [arXiv:hep-ph/9511336].
  %%CITATION = HEP-PH 9511336;%%
































\end{thebibliography}
\end{document}